\documentclass[useAMS,usenatbib]{mn2e}
\include{graphicx}
\title[An optimal Mars Trojan asteroid search strategy]{An optimal Mars Trojan asteroid search strategy}

\author[M. Todd, P. Tanga, D. M. Coward and M. G. Zadnik]{M. Todd$^{1}$\thanks{E-mail:
michael.todd@icrar.org (MT)}, P. Tanga$^{2}$, D. M. Coward$^{3}$ and M. G. Zadnik$^{1}$\\
$^{1}$Department of Imaging and Applied Physics, Bldg 301, Curtin University, Kent St, Bentley, WA 6102, Australia\\
$^{2}$Laboratoire Cassiop\'{e}e, Observatoire de la C\^{o}te d'Azur, BP 4229, 06304 Nice Cedex 04, France\\
$^{3}$School of Physics, M013, The University of Western Australia, 35 Stirling Hwy, Crawley, WA 6009, Australia}

\begin{document}

\date{}

\pagerange{\pageref{firstpage}--\pageref{lastpage}} \pubyear{}

\maketitle

\label{firstpage}

\begin{abstract}
Trojan asteroids are minor planets that share the orbit of a planet about the Sun and librate around the L4 or L5 Lagrangian points of stability. Although only three Mars Trojans have been discovered, models suggest that at least ten times this number should exist with diameters $\geq 1~km$. We derive a model that constrains optimal sky search areas and present a strategy for the most efficient use of telescope survey time that maximizes the probability of detecting Mars Trojans. We show that the \textit{Gaia} space mission could detect any Mars Trojans larger than 1~km in diameter, provided the relative motion perpendicular to \textit{Gaia's} CCD array is less than 0.40 arcsec per second.
\end{abstract}


\section{Introduction}

Trojan asteroids are minor planets that share the orbit of a planet about the Sun and librate around the L4 and L5 Lagrangian points of stability. The L4 and L5 points are $60\degr$ ahead and behind, respectively, the planet in its orbit. Trojans represent the solution to Lagrange's famous triangular problem and appear to be stable on long time-scales (100 Myr to 4.5 Gyr) \citep{pil99,sch05} in the N-body case of the Solar System. This raises the question whether the Trojans formed with the planets from the Solar nebula or were captured in the Lagrangian regions by gravitational effects. Studying the Trojans provides insight into the early evolution of the Solar System.

Since the discovery of the first Trojan in 1906 \citep{nic61} several thousand more have been found in the orbit of Jupiter. Among the terrestrial planets only Earth and Mars are known to have Trojans. While Earth has only a single known Trojan (2010 TK$_7$) which was discovered through examination of data from the WISE satellite \citep{con11}, Mars presently has three known Trojans (5261 Eureka, 1998 VF$_{31}$ and 1999 UJ$_7$) listed by the Minor Planet Center. Previous modelling \citep{tab99,tab00a, tab00b} suggests that this number represents less than a tenth of the Mars Trojan (MT) population with diameter $\geq 1~km$, and that there may be in excess of 100 with diameter $\geq 100~m$.

Non-discovery of additional MTs may be attributed to a lack of observations targeting regions in which MTs could be found, or their apparent motion being similar to inner Main Belt objects. Programmes to search for Near-Earth Asteroids (NEA) may flag objects with high apparent motion for further study while slower-moving objects are noted but may not be specifically followed up. NEA search regions are near the plane of the ecliptic and do not specifically target the entire region of stability for MTs. Observations are also limited to periods when the MT regions are visible from Earth. Other reasons for non-detection include the relatively small population, and orbital inclinations outside the plane of the ecliptic. Asteroids which could be Trojan candidates would not be flagged for further study by routine surveys because their apparent motions would not match the parameters of the survey for follow-up. 

The spatial separation between the regions allows the L5 (trailing) region to be surveyed while Mars approaches opposition, and the L4 (leading) region surveyed when Mars has passed opposition. This affords a period of several months during which these regions could each be fully surveyed. Even with such relative flexibility in the available observing period, in contrast to searching for Earth Trojans \citep{tod12b}, it is still important to find the optimal strategy for efficient use of telescope time while maximizing sky coverage and probability of detection.

This paper employs a model probability distribution which we use to constrain optimal search areas and imaging cadences for efficient
use of telescope time while maximizing the probability of detecting MTs. We examine in greater depth the case of detecting MTs from the initial study of inner planet Trojans \citep{tod12a}.

\section{Model}

Existing models \citep{tab99,tab00a, tab00b} provide estimates of MT populations, and some studies of the composition of known MTs have been made. \citet{riv03} found that (5261) Eureka and (101429) 1998 VF$_{31}$ are most likely Sa- or A-class asteroids, and that (121514) 1999 UJ$_7$ may be of X-class. 
This is an important consideration since albedo can vary greatly depending on composition. It is likely that other MTs will be relatively high albedo S-class (silicaceous) asteroids similar to NEAs and inner Main Belt asteroids, although there may be C-class (carbonaceous) asteroids. This affects the detection limit as S-class asteroids have a typical albedo of $p_{v}=0.203$, X-class asteroids have $p_{v}=0.174$ and C-class asteroids have $p_{v}=0.057$ \citep{war09}. In this paper calculations are made using these albedo values for S- and C-class asteroids to set limits on calculated magnitudes.

\begin{figure}
\includegraphics{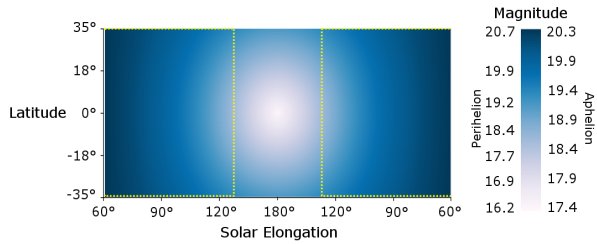}
\caption{\label{fig:figure1}Apparent magnitude of a 1~km Mars Trojan by Solar Elongation and Heliocentric Latitude. 
Brightness ranges from $V=17.4$ at opposition ($180\degr$) to $V=20.3$ at a Solar elongation of $60\degr$ when the field is at aphelion, and ranges from $V=16.2$ to $V=20.7$ when the field is at perihelion. 
Elongations $\le 135\degr$ lie within \textit{Gaia}'s scanning limit (yellow dotted line).}
\end{figure}

Calculations of absolute and apparent magnitudes at opposition using the methods described in \citet{ted05} and \citet{mor02} are shown in Table \ref{tab:table1}, neglecting atmospheric extinction.  Assuming S-class as the dominant class in the inner Solar System, and that an MT has an eccentricity similar to that of Mars, the apparent magnitude for an MT of 1~km diameter varies between $V=16.2$ at opposition to $V=20.7$ at a Solar elongation of $60\degr$ (Fig. \ref{fig:figure1}) with the field at perihelion. An MT of 100~m diameter varies in magnitude between $V=21.2$ to $V=25.7$ across this elongation range in the same fashion. If the field is at aphelion the brightness ranges from $V=17.4$ to $V=20.3$ (1-km diameter), and $V=22.4$ to $V=25.3$ (100-m diameter).

\begin{table}
 \centering

\caption{Absolute and apparent magnitudes at opposition, with apparent magnitudes for objects at perihelion and aphelion assuming eccentricity similar to that of Mars. \label{tab:table1}}

\begin{tabular}{lclccc}
\hline 
Class  & Albedo  & Diameter  & Abs. mag. & \multicolumn{2}{c}{App. mag. $(V)$}  \tabularnewline
 & & & $(H)$ & Peri. & Aph. \tabularnewline
\hline 
S-class & $0.203$ & 1.0~km & 17.35 & 16.2 & 17.4 \tabularnewline
  &   & 100~m & 22.35 & 21.2 & 22.4 \tabularnewline
\hline
C-class & $0.057$ & 1.0~km & 18.73 & 17.6 & 18.8 \tabularnewline
  &  & 100~m & 23.73 & 22.5 & 23.8 \tabularnewline
\hline 
\end{tabular}
\end{table}

\begin{figure}
\includegraphics{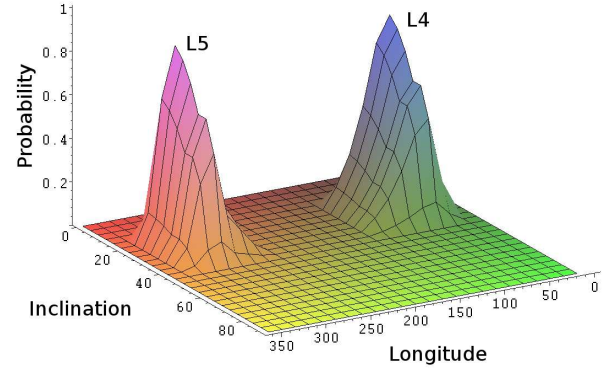}
\caption{\label{fig:figure2}Probability distribution for Mars Trojan bodies by Inclination and Heliocentric Longitude
(degrees). The figure shows peak detection probabilities for longitudes
consistent with the classical Lagrangian points but that bodies, while
co-orbital with Mars, are unlikely to be co-planar.}
\end{figure}

\citet{mik94} found that MT orbits are only stable within the inclination ranges of $15\degr \la i \la 30\degr$ and $32 \la i \la 44\degr$ over an integration period of 4~Myr. \citet{tab99} mapped stable inclinations, finding that inclinations of $15\degr \la i \la 30\degr$ were more favourable. \citet{sch05} refined this result, finding that objects with inclinations $\ga 35\degr$ become destabilized over longer periods. These orbit inclination models, and the heliocentric longitude model of \citet{tab00b}, were used to identify regions where bodies are most likely to exist (Fig. \ref{fig:figure2}). 

\begin{figure}
\includegraphics{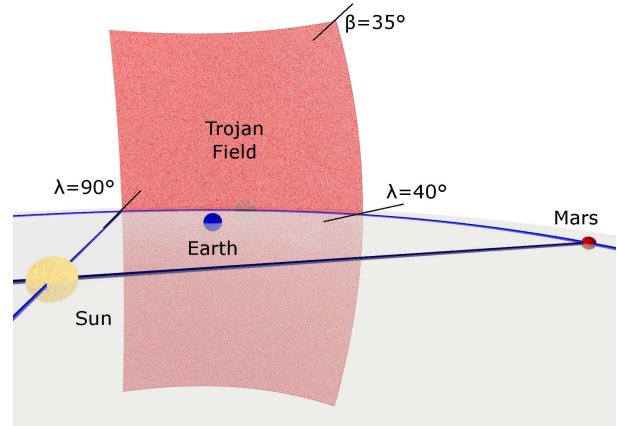}
\caption{\label{fig:figure3}Perspective illustration of Mars
Trojan (L4) target field. The field ranges from Heliocentric longitude
$(\lambda)$ $40\degr$ to $90\degr$ and latitude $(\beta)$ $-35\degr$ to $35\degr$. A complementary
field exists in the trailing Lagrangian L5 region. This illustration
represents the region in which Trojans are expected to be found, with the classical Lagrangian point at opposition.}
\end{figure}

The MT fields (Fig. \ref{fig:figure3}) are bounded by the upper inclination limit of $35\degr$ and heliocentric longitude limits (FWHM) of $40\degr \la \lambda \la 90\degr$ (L4 region) and $270\degr \la \lambda \la 320\degr$ (L5 region). About $\sim69$ per cent of projected bodies exist within these regions. We assume the distribution of bodies between the L4 and L5 regions is approximately equal.

The heliocentric solid angle of each MT field is 1.0~sr (3280~deg$^2$). Calculation of the geocentric solid angle, necessary for Earth-based observations, requires a transformation from the heliocentric reference. A numerical integration is performed using the solid angle integral described in \citet{tod12b} to determine the sky area for an Earth-based observer or a space-based instrument such as the \textit{Gaia} satellite%
\footnote{http://gaia.esa.int%
}, which will be located near the Earth's L2 Lagrangian point \citep{mig07}, for the observer's position relative to the field. 

Calculations are made for Earth's longitude corresponding to the aphelion and perihelion longitudes of Mars' orbit to determine the upper and lower limits on maximum sky area.
The calculated geocentric solid angles are 3.32~sr (10900 deg$^2$) and 5.14~sr (16900 deg$^{2}$) for the fields at opposition at aphelion and perihelion respectively. With the field centres at a Solar elongation of $60\degr$ the fields are 0.68~sr (2240~deg$^2$) and 0.62~sr (2040~deg$^{2}$) for Earth at Mars' aphelion and perihelion longitudes respectively. For \textit{Gaia}, at the L2 Lagrangian point, these values differ by less than 1.5 per cent from the values calculated for Earth. Although \textit{Gaia's} orbit prevents observing regions at opposition, it will survey elongations from $45\degr$ to $135\degr$.

\section{Telescope surveys}

The ability of current and proposed wide-field survey telescopes to survey the regions where Trojans are likely to be found in the orbits of Earth and Mars was first examined in \citet{tod12a}. It was found that the large sky area, particularly when the region is at opposition, would require a widefield telescope to survey the entire region. Attempting to complete such a survey in a single day was found to be time-consuming and inefficient. 

Table \ref{tab:table2} compares the relative capabilities of selected survey telescopes to cover the entire MT region at opposition and at a Solar elongation of $60\degr$. We show that there exists up to eight-fold difference in area, and hence time required, between the region at opposition and at $60\degr$ elongation. Although the \textit{Gaia} satellite has a narrow field-of-view (FOV), it has been included since it will operate in a continuous scanning mode and its orbit will enable it to image all of the sky to a Solar elongation of $45\degr$ \citep{mig07}. \textit{Gaia} will not be affected by the constraints of local horizon and airmass experienced by ground-based telescopes, however this advantage is mitigated by its limiting magnitude of $V=20$.

The strategy of observing a swath of the region and progressively imaging the entire field over a period of time \citep{tod12b} is suggested as an optimal method of searching for MTs.
By imaging a swath of sky between the upper and lower latitude limits, the entire field can be surveyed over time as Earth (and Mars) revolves about the Sun. A single FOV-wide swath would be imaged in minutes by a survey telescope, as shown in Table \ref{tab:table3}. Whether the traditional approach of comparing images for moving object detection or flagging uncatalogued sources, the cadence is determined by the telescope FOV. 

The relative geometry of Earth and Mars means that in a 2-year period Mars is not visible for about 3 months either side of conjunction. The same applies to the MT fields, but also means the when the L4 field is at conjunction, then the L5 field is at least partly visible. This allows the fields to be surveyed over an extended period, or at different elongations in a synodic period.
Defining specific regions and progressively observing the field over an extended period allows the entire field to be surveyed with efficient use of telescope time, and the observations accommodated around the primary science missions.

\begin{figure}
\includegraphics{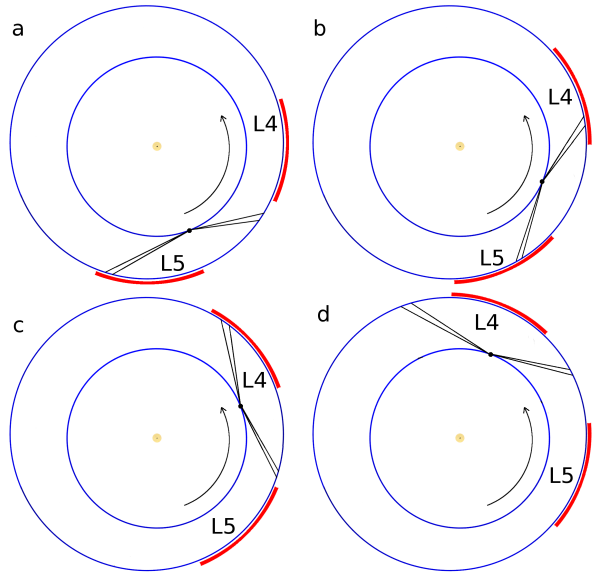}

\caption{\label{fig:figure4}Observing a defined region of sky, with Earth's orbit about the
Sun, implies the entire field can be imaged. The passage of time between each position from \ref{fig:figure4}a to \ref{fig:figure4}d is about 1.5 months.}

\end{figure}

\begin{figure}
\includegraphics{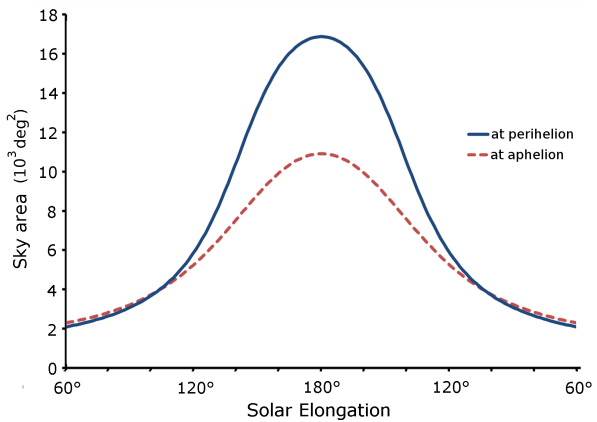}

\caption{\label{fig:figure5}The sky area of each region varies from $\sim2000$~deg$^{2}$ at a Solar elongation of $60\degr$ to 11000--17000~deg$^2$ at opposition ($180\degr$). The change in distance between Earth's and Mars' orbit, primarily due to Mars' eccentricity, significantly affects the sky area when the field is near opposition.}

\end{figure}

\begin{figure}
\includegraphics{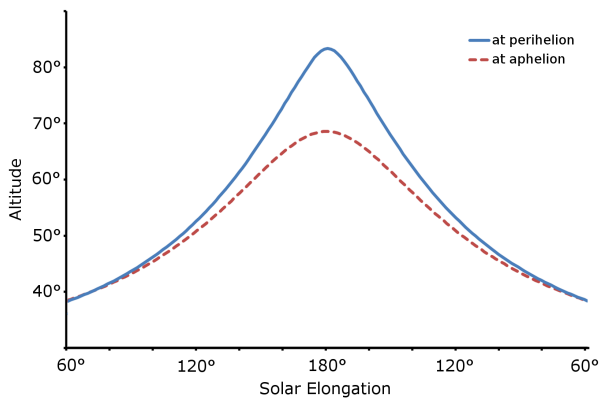}

\caption{\label{fig:figure6}The angular height of the Mars Trojan field above (and below) the ecliptic plane from the geocentre at opposition ranges from $66\degr$ at aphelion, or $83\degr$ at perihelion, to $\sim 30\degr$ at a Solar elongation of $60\degr$.}

\end{figure}

\begin{table*}
 \centering
 \begin{minipage}{180mm}

\caption{Comparison of different survey telescopes showing the required time
to survey the entire field at opposition and at a Solar elongation of $60\degr$. At opposition the sky area ranges from 10900~deg$^2$ (aphelion) to 16900~deg$^2$ (perihelion) while at an elongation of $60\degr$ it ranges from 2240~deg$^2$ (aphelion) to 2040~deg$^2$ (perihelion).\label{tab:table2}}

\begin{tabular}{llccccccl}
\hline 
Telescope  & Limiting mag.  & Exposure  & FOV  & \multicolumn{2}{c}{Opposition} & \multicolumn{2}{c}{$60\degr$ elongation} & Instrument capabilities \tabularnewline
 &  &  &  & (FOVs) & (Time) & (FOVs) & (Time) &  \tabularnewline
\hline 
Catalina  & $V\sim20$ & 30~s & 8.0~deg$^2$ & 1360--2110 & 11.3h--17.6h & 280--255 & 2.3h--2.1h & \citep{dra09} \tabularnewline
PTF 1.2~m  & $R\sim20.6$  & 60~s  & 8.1~deg$^2$ & 1345--2085 & 22.4h--34.8h & 277--252 & 4.6h--4.2h & \citep{law09} \tabularnewline
Pan-STARRS & $R\sim24$ & 30~s & 7.0~deg$^2$ & 1555--2415 & 13.0h--20.1h & 320--292 & 2.7h--2.4h & \citep{jed07} \tabularnewline
LSST  & $r\sim24.7$ & 30~s & 9.6~deg$^2$ & 1135--1760 & 9.5h--14.7h & 234--213 & 2.0h--1.8h & \citep{jon09} \tabularnewline
Gaia & $V\sim20$ & 39.6~s$^{\dagger}$ & 0.69\degr-wide & 160--195$^{\ddagger}$ &  & 55--50$^{\ddagger}$ & 330h--300h$^{\ast}$ & \citep{mig07} \tabularnewline
\hline 
\end{tabular}

$\dagger$\textit{Gaia} will operate in a continuous scanning mode where the CCD array will be read out at a rate corresponding to the angular rotation rate of the satellite (6h period).

$\ddagger$\textit{Gaia}'s orbit parameters prevent observations within $45\degr$ of opposition. These values represent how many rotations it would take for \textit{Gaia} to survey a region of this size. \textit{Gaia}'s specific precession parameters are not considered so values should be considered as representative.

$^{\ast}$Approximate time to complete sufficient rotations to scan across the entire field. The actual time spent scanning within the field will be a fraction of this value. 

\end{minipage}
\end{table*}

\begin{table*}
 \centering
 \begin{minipage}{180mm}

\caption{Comparison of different survey telescopes showing the required time
to survey a single swath of one CCD- or detector-width of the field at opposition compared to a Solar elongation of $60\degr$. At opposition the height of the field ranges from $66\degr$ (aphelion) to $83\degr$ (perihelion) above the ecliptic plane while at an elongation of $60\degr$ the height is $\sim 30\degr$.\label{tab:table3}}

\begin{tabular}{lccccccl}
\hline 
Telescope  & \multicolumn{2}{c}{Opposition} & \multicolumn{2}{c}{$120\degr$ elongation} & \multicolumn{2}{c}{$60\degr$ elongation}  \tabularnewline
 & (FOVs) & (Time) & (FOVs) & (Time) & (FOVs) & (Time) &  \tabularnewline
\hline 
Catalina  & 47--59 & 23.5--29.5 min & 37--46 & 18.5--23 min & 21--22 & 10.5--11 min  \tabularnewline
PTF 1.2~m  & 47--59 & 47--59 min & 37--46 & 37--46 min & 21--22 & 21--22 min  \tabularnewline
Pan-STARRS & 50--63 & 25--31.5 min & 40--49 & 20--24.5 min & 22--23 & 11--11.5 min  \tabularnewline
LSST  & 43--54 & 21.5--27 min & 34--42 & 17--21 min & 19--20 & 9.5--10 min \tabularnewline
Gaia$^{\dagger}$ &  & 131--166 min$^{\ddagger}$ & & 104--130 min & & 60--61 min & \tabularnewline
\hline 
\end{tabular}

$\dagger$\textit{Gaia} will operate in a continuous scanning mode where the CCD array will be read out at a rate corresponding to the angular rotation rate of the satellite (6h period). \textit{Gaia}'s specific precession parameters are not considered so values should be considered as representative.

$\ddagger$\textit{Gaia}'s orbit parameters prevent observations within $45\degr$ of opposition. 

\end{minipage}
\end{table*}

On approach to the trailing edge of the L5 MT region, observations can be made in the morning before sunrise. Thus end-of-night observations could be set to image across L5 on approach. When Mars is near opposition, morning observations can be made of the L4 region and evening observations can be set up to re-cover the L5 region. After the L4 region has passed opposition evening observations can be made to re-cover that region. Morning observations will have ceased. The progression in Figure \ref{fig:figure4} shows how the MT regions might be surveyed as Earth passes by in its orbit.

Observations of MTs are time-limited only by the relative orbits of Earth and Mars. There will exist specific constraints particular to the geographic location of a telescope, depending on the relative positions of Earth and the MT field. For example some Northern Hemisphere telescopes may not have access to the entire Southern part of the MT region described in Figure \ref{fig:figure3} when the field is near opposition, and the converse will apply to some Southern Hemisphere telescopes. 

The limit adopted in this paper is that of Solar elongation of $60\degr$, but with the MT regions each being visible for several months as Earth passes by, it is possible to survey the fields at a wide range of Solar elongations. If the observations are made at small elongations the available time is limited whereas at (or near) opposition the amount of sky area can be the limiting factor. Figure \ref{fig:figure5} shows the difference in sky area of the field between opposition and $60\degr$ elongation.

While a delay between follow-up images introduces other variations from such things as changes in atmospheric conditions and seeing, this could be compensated for by image convolution. Some telescopes are implementing image processing systems designed specifically for asteroid detection (e.g. Pan-STARRS+MOPS -- Moving Object Processing System) \citep{jed07}. Depending on the relative positions of Earth and the MT field the apparent motion will vary with distance and direction.

\textit{Gaia's} orbit dictates that it will scan narrow strips of the sky. As it will be a scanning instrument that performs a continuous scan by smooth and regular rotation of the spacecraft, follow-up observations for any object must be made by another instrument. Consequently a network of ground-based telescopes, the \textit{Gaia} Follow Up Network for Solar System Objects (Gaia-FUN-SSO)%
\footnote{http://www.imcce.fr/gaia-fun-sso/%
} is being established to provide follow-up observations after detection of Solar System objects. 

The scanning law governing \textit{Gaia's} operation and the way the CCDs will operate \citep{tan08,mig10} limits the detection of Solar System objects as they cross the CCD array. With the $106.5\degr$ separation between lines of sight, motion in the `across-scan' (AC) direction of 0.40 arcsec per second will cause an object to traverse one FOV-width and so have passed out of the FOV between the first and second observations of the field, even if starting at the edge of the CCD array on the first pass. An AC motion greater than 0.040 arcsec per second is larger than the pixel size in the AC direction and will cause smearing across pixels. Similarly, an `along-scan' (AL) motion greater than 0.013 arcsec per second is larger than the pixel size in the AL direction. Since the motion of a Solar System object across the CCD depends on the relative positions of the object and \textit{Gaia}, and the particular scanning orientation of \textit{Gaia} at that time, these are important considerations for the design of the algorithms for detection of Solar System objects.

\section{Summary}

Despite the thousands of known Jupiter Trojans, a mere handful of terrestrial Trojans have been discovered. There are only three known Mars Trojans, and the first Earth Trojan has only been very recently discovered \citep{con11}. Simulations \citep{tab99, tab00a, tab00b} predict that this number constitutes about a tenth of the MT population with diameter $\geq~1~km$. The prospect of detecting this population is limited by the size of the stable regions in which MTs can exist. However the Earth-Mars geometry offers an extended period over which a survey could be conducted. The conventional method of detecting asteroids by repeated observations of a field can be used, with a systematic approach to survey the entire MT field during a synodic period.

This study has identified the region of highest probability for detection, with an inclination $\leq 35\degr$ and heliocentric longitude range of $40\degr$ -- $90\degr$ (L4) and $270\degr$ -- $320\degr$ (L5). Surveys of the entire field within the chosen limits are impractical on telescopes with small FOV but are possible on survey telescopes with sufficient FOV to accomplish the task in a single night, such as Catalina or the Large Synoptic Survey Telescope (LSST)%
\footnote{The LSST is still in the development phase (www.lsst.org).%
}, when the field is at a Solar elongation $\la 150\degr$. In these cases it may be possible to survey the regions defined by the inclination limits. This would require the fields to be surveyed twice within a few days, as is common for Main Belt asteroids.  Although possible, it is rather impractical as this occupies a significant amount of telescope time. 
When the field is at a Solar elongation $> 150\degr$ the sky area becomes too large to be able to survey in a single night by any existing widefield telescope.

Given the challenges involved in attempting to survey the entire field in a single session, we aimed to minimise the time requirement. Observing a swath of sky each session and progressively sampling the entire field over a synodic period achieves this aim. This approach requires minimal time each session by making a single pass across the field. By repeating this at intervals which provides an overlap of the telescope FOV from one pass to the next the common regions of consecutive images can be compared. 

The sky areas for Earth-based observers at different Solar elongations have been determined using the numerical integrations described in \citet{tod12b}. A strategy has been proposed for observing a sub-region of the MT field and, as Earth revolves about the Sun, redefining this region and progressively surveying the entire field during a synodic period. This approach takes only a few minutes per night at intervals of 3--4 days. This approach requires the observed region to be redefined at intervals to progressively observe the entire field over those months when the field is visible from Earth. 

Since the field is visible during the majority of one synodic period (about two years), nights lost due to adverse weather conditions can be rescheduled without critically impacting the timing of the programme. The flexibility in the timing allows such a programme to be more readily accommodated alongside the primary science mission, and is readily achievable by a survey telescope. While this method requires a program of continued observations over several months, the total time commitment for the program is a few tens of hours spread over that period. 

The \textit{Gaia} satellite, at Earth's L2 Lagrangian point, will survey the MT fields as part of its larger mission to survey the whole sky. With a detection limit of $V=20$, \textit{Gaia} can be expected to detect MTs larger than 1~km diameter, provided the lateral motion across the CCD array is less than 0.40 arcsec per second. As with any uncatalogued Solar System object detected by \textit{Gaia}, these would need to be followed up by ground-based telescopes. 

The specific observing geometry of the \textit{Gaia} satellite at Earth's L2 Lagrangian point will be examined in more detail in future work. Initial simulations for \textit{Gaia's} detection of inner Solar System Trojans in the orbits of Earth and Mars show promise. Results of detailed simulations will be reported with particular regard to the detection limits and observational mode of operation of \textit{Gaia}.

\section*{Acknowledgments}
MT thanks the organizers of the \textit{Gaia} Solar System Science workshop (held in Pisa, Italy, 2011) for providing a fertile environment for discussing \textit{Gaia} science. DMC is supported by an Australian Research Council Future Fellowship.

\label{lastpage}

\end{document}